\documentclass{article}

\usepackage{graphicx}
\usepackage{psfig}
\usepackage{epsfig}
\usepackage[round]{natbib}

\title{Five guidelines for the evaluation of site-specific medium range probabilistic temperature forecasts}

\author{Stephen Jewson and Christine Ziehmann
\footnote{\emph{Correspondence address}: RMS, 10 Eastcheap,
London, EC3M 1AJ, UK. Email: \texttt{x@stephenjewson.com}}\\
Risk Management Solutions, London, United Kingdom}

\begin{document}

\maketitle

\begin{abstract}

Probabilistic temperature forecasts are potentially useful to the energy and weather derivatives industries.
However, at present, they are little used. There are a number of reasons for this, but we 
believe this is in part due to inadequacies in the methodologies that have been used to evaluate such forecasts,
leading to uncertainty as to whether the forecasts are really useful or not and making it hard
to work out which forecasts are best.
To remedy this situation we describe a set of guidelines that we recommend should be followed when evaluating the 
skill of site-specific probabilistic medium range temperature forecasts. 
If these guidelines are followed then the results of validation can be used directly by forecast
users to make decisions about which forecasts to use. 
If they are not followed then the results of validation may be \emph{interesting}, but will not
be practically useful for users.
We find that none of the published studies that evaluate such forecasts fall within our guidelines, 
and that, as a result, none convey the information that the users need
to make appropriate decisions about which forecasts are best.

\end{abstract}

\section{Introduction}

Meteorological forecasts attempt to convey information about uncertain future weather states.
They can consist of single values, or of a distribution of values with probabilities for each outcome.
In the single value case, the value should usually be interpreted as the mean of the distribution
of future outcomes. In the probabilistic case, the probabilities are hopefully good indications
of the real probabilities of possible events.

Both forecasts of the mean and probabilistic forecasts have potential applications in 
the energy and weather derivatives industries: our examples will come from the latter.
Forecasts of the mean are useful when one wishes to evaluate the expectation of a linear function
of the temperature. Calculating the fair strike for a linear weather swap contract is an example
of this (see~\citet{jewsonz03a} for details). 
Probabilistic forecasts are useful in two ways: firstly, when one wishes to evaluate the 
expectation of a non-linear function of the temperature, 
and secondly, when one wishes to evaluate the distribution of values that can 
be attained by any function of the temperature. 
Calculating the fair premium for a weather option
contract is an example of the first case, and 
calculating the distribution of outcomes for any weather
derivative is an example of the second
(see~\citet{jewsonc03b} for details).

Forecasts for the mean temperature are usually produced using linear regression applied to the
output from numerical weather prediction models. Traditionally the input for the regression has
been the output from a single integration of such a model. More recently the input for the regression 
has been taken from
the mean of an ensemble of such integrations. All these regressions typically deal with 
temperature that has been converted into
either
single anomalies (from which the seasonal cycle of the mean has been removed) or double
anomalies (from which the seasonal cycle of the mean and variance has been removed). 

Forecasts for the distribution of temperature are most easily produced as a by-product of the regression
step used to produce the forecast for the mean temperature. Standard regression routines 
fit the variance of the residuals, which can be taken as the variance of the predicted temperature distribution.
Probabilistic forecasts can also be produced from the distribution
of the members of an ensemble forecast. The hope is that such an ensemble based method will give better
probabilistic forecasts than the regression based approach because it has the potential to give forecasts in 
which the uncertainty varies with the state of the atmosphere. However, at this point in time, little 
research has been done to compare these two methods for generating probabilistic forecasts, 
and to assess whether the ensemble member
based approach really can give better predictions. Results in~\citet{jewson03g}
suggest that probabilistic forecasts based on regression on the ensemble mean are actually very hard to beat.

Users of meteorological forecasts in the energy and weather derivatives industries are 
well able to understand the potential value of good probabilistic forecasts. The pricing of many types of
energy and weather derivative contracts are based on a fully probabilistic analysis of possible future outcomes,
and incorporating probabilistic forecasts into these pricing algorithms is not necessarily very difficult.
However, probabilistic forecasts are at present rather little used. 
There are a number of reasons for this, and we will not try to identify them all here. 
Rather, we will focus on one
particular reason which is to do with the methodologies used to
verify and present probabilistic forecasts. 

Users of forecasts are generally skeptical, not to say cynical, about the claims of meteorological forecasters. This is, 
perhaps, justifiable. There is a large measure of moral hazard involved in believing 
in the analysis of a forecast when it is performed by people who have something to gain from the success of that forecast. 
For this reason, it is essential that analyses of forecasts should be performed with strictly correct methodologies
that clearly address the questions that users need to see addressed.
However, a quick perusal of the academic literature on the use of probabilistic forecasts 
suggests that this is not the case. 
We will not review this literature in detail: suffice to say that we have not been able to find a single paper
(including our own) that gives useful information about real site-specific forecasts~\footnote{if you have written
one, please get in touch!}.
Many of the papers we have read 
(such as~\citet{roulstons03}, \citet{taylorb}, \citet{jewsonbz03a})
get many of the
methodologies correct, and produce \emph{interesting} results, 
but none of them go quite far enough.
Unfortunately, in order to prove that a forecast
is \emph{truly} useful, rather than just \emph{possibly} useful, one has to get the whole methodology correct.
Consequently, it would seem impossible, from the published literature alone, to work out whether the available
probabilistic forecasts really have any practical use, and to work out how to use them.
There are plenty of good ideas, and some strong indications
that such forecasts \emph{may} be useful, but none of the articles really gets to the heart of the matter and
show that they \emph{really are} useful.

In order to contribute to the goal, of, hopefully, being able to show that probabilistic forecasts are useful,
we list below five guidelines which we believe are necessary, and perhaps even sufficient, conditions
for proving that a probabilistic forecast is useful in practice.

\subsection{Skill measures}

This article is not about skill measures. However, throughout there is an implicit assumption that some
skill measure is available which allows us to test whether one probabilistic forecast is better than another.
At the time of writing, we ourselves prefer likelihood-based skill measures (see~\citet{jewson03d}).

\section{The Guidelines}

\textbf{1. Comparison with observations not analysis}
\\

Our first guideline is that forecasts should be compared with real ground-based observations, rather
than analyses. Comparison with analyses is routinely used by forecasting agencies as a way of
evaluating forecasts, mainly because it is convenient, and allows comparison of upper-air and gridded fields.
However, the results of comparison with analysis are not useful when trying to show that forecasts
of ground-based observations have any skill: they can only show the potential that such
skill might be possible. There is, however, many a slip twixt analysis and observation and only a 
proper comparison with ground-based observations really answers the questions that users of forecasts 
need to see answered.
\\

\textbf{2. Use of skill measures not correlations}
\\

Correlations between forecasts and observations, and between the spread of ensemble forecasts and the skill, 
are often used as measures of the potential skill in a forecast. But they do not necessarily
translate into real skill, and so, in the end, cannot be used to indicate that a forecast is useful, only 
that it \emph{might} be useful.
\\

\textbf{3. Comparison with appropriate simple models}
\\

There are a number of situations in which forecasts should be compared with, and beat, forecasts from an appropriate
simpler model before one can claim the forecast is useful.

\begin{itemize}

\item
In order to prove that a forecast of the mean temperature derived from an ensemble forecast is useful 
one has to compare with the best forecast of the mean temperature that can be derived from a non-ensemble 
forecast.

\item
In order to prove that a probabilistic forecast derived from the distribution of the members of an ensemble is useful
one has to compare with the best probabilistic forecast that can be derived from the ensemble mean alone
using past forecast error statistics.

\item
In order to prove that a probabilistic forecast derived from the ensemble mean is useful
one has to compare with the best probabilistic forecast that can be derived from a single member of the
ensemble using past forecast error statistics.

\item
It is \emph{not} useful, however, to compare a probabilistic forecast from an ensemble with a single forecast from
a single model integration as such a comparison confuses two issues: whether probabilistic forecasts are
better than single forecasts, and whether ensembles are a useful way to make probabilistic forecasts.

\item
It is \emph{not} useful to compare probabilities derived from the distribution of the members of an ensemble with
probabilities derived from a single model integration and past forecast error statistics because this also confuses
two issues: whether the distribution of the members of the ensemble contains useful information, and whether
the ensemble mean contains useful information.

\end{itemize}

In those cases where past forecast error statistics are used to create a probabilistic forecast, the simplest
model to use is linear regression.
\\

\textbf{4. In sample and out of sample tests}
\\

In-sample tests can be useful in the sense that if forecast A, derived from a complex calibration model, does not beat
forecast B, from a simple calibration model, then it is very unlikely to beat it out of sample. However, the reverse is not
true: if forecast A, from a complex calibration model, 
\emph{does} beat forecast B, from a simple calibration model, in in-sample tests, this
usually proves nothing about what will happen out of sample. For this reason, out of sample tests are always needed
if one is attempting to prove that a more complex calibration model is better.
\\

\textbf{5. Avoiding aggregation}
\\

It is common practice when validating forecasts to aggregate temperatures over a region.
Because the information in forecasts is greater on larger scales, this usually gives
better validation results. There may be cases where users of forecasts \emph{are} interested in forecasts for large
scales, but in most cases forecast users are specifically interested
in individual sites. This is particularly the case for the weather derivative market. 

It is also common practice to aggregate the \emph{results} of validation over many locations
(as opposed to aggregating the temperature before validation) 
especially amongst national met services who may produce
forecasts for hundreds of locations. This may be useful as an overall performance measure,
but it is not useful for users of forecasts who are typically interested in performance at individual sites.

\section{Summary}

Many papers and reports have been written which show that certain probabilistic forecasts \emph{might} be useful.
We have presented five methodological guidelines that should be followed if one wishes to show
that a forecast really \emph{is} useful.

\newpage
\bibliographystyle{plainnat}
\bibliography{guidelines}

\end{document}